\documentclass{webofc}
\usepackage[varg]{txfonts} 

\usepackage{array}
\usepackage{multirow}
\usepackage[frozencache,cachedir=.]{minted}
\usepackage{hyperref}
\usepackage{natbib}

\hypersetup{
    colorlinks=true,
    urlcolor=blue
}

\begin{document}

\title{FuncADL: Functional Analysis Description Language}

\author{
    \firstname{Mason} \lastname{Proffitt}\inst{1}\fnsep\thanks{\email{masonLp@uw.edu}} \and
    \firstname{Gordon} \lastname{Watts}\inst{1}\fnsep\thanks{\email{gwatts@uw.edu}}
}

\institute{University of Washington}

\abstract{
    The traditional approach in HEP analysis software is to loop over every event and every object via the ROOT framework.
    This method follows an imperative paradigm, in which the code is tied to the storage format and steps of execution.
    A more desirable strategy would be to implement a declarative language, such that the storage medium and execution are not included in the abstraction model.
    This will become increasingly important to managing the large dataset collected by the LHC and the HL-LHC.
    A new analysis description language (ADL) inspired by functional programming, FuncADL, was developed using Python as a host language.           
    The expressiveness of this language was tested by implementing example analysis tasks designed to benchmark the functionality of ADLs.
    Many simple selections are expressible in a declarative way with FuncADL, which can be used as an interface to retrieve filtered data.
    Some limitations were identified, but the design of the language allows for future extensions to add missing features.
    FuncADL is part of a suite of analysis software tools being developed by the Institute for Research and Innovation in Software for High Energy Physics (IRIS-HEP).
    These tools will be available to develop highly scalable physics analyses for the LHC.
}

\maketitle

\section{Introduction}
\label{intro}

Analyses of high energy physics (HEP) data typically consist of executing a carefully designed algorithm on millions of collision events.
The algorithm is specific to the analysis but identical for all events in a given data-taking period.
The per-event results are aggregated over the entire dataset, such as in the form of histograms or a table recording the number of events passing certain selections.
One of the most common ways to write analysis code is essentially a for-loop over events, generally using C++ and the ROOT software framework \cite{ttree, eventloop}.
This procedure tightly couples the analysis code to both the file format of the data and the steps of program execution.

There have been several recent efforts to provide more declarative interfaces for data analysis in order to simplify analysis code and to allow taking advantage of optimization via parallelization and vectorized operations.
Within the ROOT package, the \texttt{RDataFrame} class aims to provide these features by drawing on the concept of data frames present in other platforms like R, Apache Spark, and the Python package pandas \cite{rdf}.
Coffea  is a CMS-focused project designed to handle cuts and histograms similarly to \texttt{RDataFrame}, but entirely in Python, utilizing the Uproot package for reading and writing ROOT files \cite{coffea}.

Many of the issues faced in HEP data analysis are similar to those mitigated by work in database management systems.
Two important aspects of database management are data query languages and data independence \cite{textbook}.
Query languages such as SQL are naturally declarative, as they only specify relationships between data fields.
Data independence is the concept that the interface for accessing data should not depend on how and where the data is actually stored.
There is not yet a widely adopted declarative interface to write HEP analyses across a variety of underlying storage formats.
The object of the FuncADL project is to provide this interface via a query language that maintains data independence.

The FuncADL project encompasses several different aspects: the query language, code generation, and implementations that tie these two steps together.
The following sections address these components.
A simple implementation is demonstrated with example analysis tasks drafted by the HEP software community, and there is a discussion of implications and future plans.

\section{Query language}
\label{language}

FuncADL has been designed as a query language within Python \cite{funcadl}.
The fundamental base class is \texttt{EventDataset}, which generically represents a dataset of collision events, regardless of the storage format or location.
All queries are constructed by applying operations on an instance of \texttt{EventDataset}.
The query operations were inspired by LINQ (Language INtegrated Query), a feature built into the C\# language \cite{linq}.
Query operators always act on a sequence of objects, which may be the sequence of events or a sequence that exists within each event, such as a jet collection.
The most important query operations can be categorized as projection, filtering, or aggregation operators.

Projection operators apply a transformation to each element of a sequence.
This transformation is based on the properties of each element.
For example, this can be used to select only the properties of an event relevant for a particular analysis and to disregard the rest of the information.
\texttt{EventDataset} itself is effectively a trivial projection operator passing all event information through.
There are two non-trivial projection operators: \texttt{Select} and \texttt{SelectMany}.
Both of these take a lambda function as an argument that specifies the transformation to apply to each sequence element.
The result of \texttt{Select} is simply the sequence of results of this transformation.
The \texttt{SelectMany} projection operator expects this transformation to produce a sequence for each element, and the operator concatenates all of these sequences.
This allows flattening collections across all events, such as all jets in a dataset.

Filtering operations filter out elements of a sequence based on a certain condition.
The filtering operator is \texttt{Where}.
\texttt{Where} expects a lambda function argument that results in a boolean value.
An element of the input sequence only passes through the filter if applying the lambda function to it evaluates to \texttt{True}.
Otherwise the sequence element is dropped.

Aggregation operators map an input sequence to a single value.
These act on each element sequentially and reduce the dimensionality.
The generic form is \texttt{Aggregate} operation, which takes a lambda function and a seed value as arguments.
The lambda function is executed on the first element of the sequence with the seed value as a second argument.
The resulting value becomes the seed value for applying the function to the second element, which provides the seed value for the third element, and so on.
There are several common aggregation operations that are specializations of the \texttt{Aggregate} operator which are implemented directly in the language for convenience.
These include the operators \texttt{Count}, \texttt{Max}, \texttt{Min}, and \texttt{Sum}.
\texttt{Count} outputs the number of elements in the input sequence.
\texttt{Max}, \texttt{Min}, and \texttt{Sum} expect the input sequence to be numerical values and output the maximum element, the minimum element, and the sum over all elements, respectively.

Only an example subset of operators have been enumerated here.
There are other operations supported for more specialized usage.
In addition, several native Python types and operations retain identical functionality within FuncADL.
Lists, tuples, and dictionaries are allowed within queries.
All arithmetical, boolean, and bitwise Python operators are supported as well.
One important special operation supported in FuncADL is \texttt{Zip}.
\texttt{Zip} takes several sequences of the same length and turns them into one sequence by bundling together the elements at a given position in each input sequence.
The functionality is similar to the Python builtin class \texttt{zip} \cite{zip}.
However, rather than only yielding tuples, FuncADL's \texttt{Zip} operator also allows zipping dictionaries together to output a sequence of dictionaries.
This makes it possible to create new ad hoc data structures inside a query.

Useful queries are generally composed of several operations.
These can be applied in succession on the same sequence, or they can be nested such that some of the query operations are run on sequences within each event.
The level of nested queries can continue to any depth of object structure present in the data.

\section{Code generation}
\label{code-gen}

The key to achieving data independence with a query language is having backends that can handle the details of how to actually execute a query. 
For FuncADL, this means taking a representation of the user's query and producing a script or executable that will apply the required transformations to the data.
Currently two FuncADL backend libraries exist: one utilising the \mbox{ATLAS} AnalysisBase software \cite{athena} and one using Uproot \cite{uproot}.
These backends are designed to run on xAOD \cite{xaod} files and flat ROOT ntuples, respectively.
Here a "flat ntuple" refers to a ROOT \texttt{TTree} that does not require any external schema to interpret its data structures.
For example, this includes the CMS NanoAOD \cite{nanoaod} format and potentially the developing \mbox{ATLAS} DAOD\_PHYSLITE \cite{physlite} format, as well as most ntuples produced by physics analysis teams.

The FuncADL query is internally represented by an abstract syntax tree (AST).
An AST is composed of nodes corresponding to syntactic elements of Python and FuncADL, such as a lambda function or a \texttt{Select} operation.
Each node contains links to child nodes, corresponding to the components of that syntactic element (for example, the arguments of a lambda function and the body of the function).
The AST is built using the \texttt{ast} module included in Python \cite{ast}.
The backend implementations of FuncADL take the AST of a query and traverse the tree with an \texttt{ast.NodeTransformer} (or an \texttt{ast.NodeVisitor}).
For each node, these \texttt{NodeTransformer}s build a representation of the corresponding operation in the appropriate framework for the particular backend.
This process translates a query AST into standalone code that will perform the retrieval and transformations specified by the query.
The output is the generated code text, which can be compiled or executed as needed.

One of the currently available backend implementations is for the xAOD format.
This is the format used by ATLAS for data files produced by the collaboration for general use by analysis teams.
The xAOD format is a specially structured ROOT file produced by the Athena software framework.
Because of the data structures used, it is generally necessary to run a collaboration-specific framework like AnalysisBase in order to usefully read and extract information from the files.
Thus a FuncADL backend for xAOD files was written that generates C++ code utilizing the ATLAS xAOD libraries.
This allows accessing xAOD data structures from within the FuncADL query language, such as leptons and jet collections.

Another backend available for FuncADL is based on Uproot and Awkward.
Uproot is a Python package completely independent from any ROOT code that is able to read ROOT files.
Awkward is a related Python package that provides an interface for accessing and manipulating arrays with potentially jagged dimensions \cite{awkward}.
These packages are designed to be lightweight and fast, benefiting from compiled vectorized optimizations.
Uproot is capable of reading any ROOT \texttt{TTree}s that contain standard data types like strings, integers, floating point values, and variable-length vectors.
This makes it ideal for operating on ROOT files with a simple structure like NanoAOD (CMS) or DAOD\_PHYSLITE (ATLAS).
The Uproot FuncADL backend translates a query AST into generated Python source code for a function that can be evaluated on a data file and returns the selected and transformed values as an Awkward array.

New backends for FuncADL can be implemented at any time.
All that is required is to write a new \texttt{NodeTransformer} that can take in a FuncADL query AST and produce a script or executable which applies the necessary transformations to a data file of the appropriate format.
It is possible to have multiple backends that can operate on the same format.
For example, a prototype for an \texttt{RDataFrame}-based backend was created for an early version of FuncADL.
The Uproot and \texttt{RDataFrame} backends are both able to run on flat ntuples, so this provides the option for alternative backends and performance benchmark comparisons between them.

\section{Implementations}
\label{implementations}

The final piece of implementing FuncADL is tying the frontend query language to the backends via an interface that is able to execute a query and return the result to the user.
There are a few different options for this, depending on where the query should actually be executed.
The main options available are running via a suite of services called ServiceX \cite{servicex} or running the full transformation locally in the same process that is submitting the query.
The choice is determined by which subclass of \texttt{EventDataset} is used as the base for the FuncADL query.
In both cases, the communication between the frontend parsing the user's query and the backend generating and executing code is done via a format called Qastle \cite{qastle}.
The query built by combining an \texttt{EventDataset} instance and operations acting on it is only evaluated on demand.
That is, the query is only executed when the \texttt{.value()} method is called.
Therefore the construction of the query can be done in multiple lines and can be separated from the evaluation step.

Query AST Language Expressions (Qastle) is a way of specifying data queries detached from any host language.
Qastle is essentially a plain text format consisting of LISP-like \cite{lisp} expressions corresponding to the nodes of a query AST.
This format was created to provide a human-readable interface between the frontend and the backends.
The Qastle format also removes extraneous information from the generic Python AST structure that is not relevant to FuncADL queries.
The FuncADL frontend translates a user query into Qastle and passes it to the executor associated with the particular \texttt{EventDataset} subclass used.

ServiceX is a full-featured data delivery service.
ServiceX itself consists of several services that perform dataset resolution, code generation, and data transformation.
It is possible to run ServiceX locally, although it is intended to be run on a cluster as a highly scalable platform.
Query execution via ServiceX is possible through the classes \texttt{ServiceXSourceXAOD} and \texttt{ServiceXSourceUpROOT} for the xAOD and Uproot backends, respectively.
The \mbox{FuncADL} frontend package produces a Qastle-formatted query and sends this to ServiceX.
The FuncADL backends drive the code generation and data transformation steps.
The result is then sent back to the analysis user.

It is also possible to run FuncADL queries outside of ServiceX.
For example, the \texttt{UprootDataset} class provides a way to execute queries using the FuncADL Uproot backend.
With this method, the code generated for a query by the backend is simply run locally after calling the \texttt{.value()} function on the query.
This functionality is useful for testing and for applying transformations to small datasets that don't necessitate the use of cluster-scale resources.

The modular design of the frontend and backends of FuncADL allows for the possibility of adding further methods of execution in the future.
It is possible to drop in a replacement for the FuncADL frontend as well.
Any package that is capable of producing the Qastle format can be used in conjunction with the backends.
This means that it is possible to run queries with the FuncADL backends that were not even written in the FuncADL query language.
This strategy has been implemented by the package \texttt{tcut\_to\_qastle} \cite{tcut}, which translates ROOT \texttt{TCut}-formatted strings into Qastle, and these \texttt{TCut} queries can then be executed via ServiceX.

\section{Query examples}
\label{examples}

In order to craft concrete examples of FuncADL queries, a list of eight ADL benchmark tasks were used \cite{adlbenchmarksindex}.
This list was inspired by conversations within the HEP Software Foundation Data Analysis Working Group on typical simple data transformations needed in writing an analysis.
The tasks are performed on CMS open data that has been converted to the NanoAOD format.
For simplicity and because the example dataset was relatively small (16 GiB), these queries were run using local execution.
They were run in Jupyter notebooks via CERN's SWAN service \cite{swan}.

The following two setup lines are common to all of the benchmark task implementations:
\begin{minted}{python}
from func_adl_uproot import UprootDataset
ds = UprootDataset('Run2012B_SingleMu.root')
\end{minted}
The argument to \texttt{UprootDataset} is the path of the data file.
The selections in the first four benchmark tasks have been fully implemented in FuncADL, as shown in Table \ref{table:implemented}.
Only the FuncADL queries are shown here, but the resulting histograms made from the returned arrays can be seen in the GitHub repository \cite{funcadlbenchmarks}.
The last four benchmark tasks could not be implemented directly in FuncADL because of limitations in the current version, as shown in Table \ref{table:unimplemented}.
The missing features are a cross join (or cartesian product), sequence sorting, and the calculation of invariant mass.
However, it is possible to apply partial selections with \mbox{FuncADL} and implement the remaining aspects of each benchmark task by using the \mbox{FuncADL} selection output in combination with other Python packages.

\begin{centering}
\begin{table}[ht]
\frenchspacing
\renewcommand{\arraystretch}{1.3}
\begin{tabular}{|>{\centering\arraybackslash}m{2mm}|m{\textwidth-11mm}|}
\hline
\# & ADL benchmark task and corresponding FuncADL query \\
\hline\hline
\multirow{2}{*}{1} & Plot the missing $E_\textrm{T}$ of all events. \\
\cline{2-2}
&
\begin{minted}{python}
ds.Select(lambda event: event.MET_pt)
\end{minted}
\\
\hline
\multirow{2}{*}{2} & Plot $p_\textrm{T}$ of all jets in all events. \\
\cline{2-2}
& \begin{minted}{python}
ds.SelectMany(lambda event: event.Jet_pt)
\end{minted}
\\
\hline
\multirow{2}{*}{3} & Plot $p_\textrm{T}$ of jets with $|\eta| < 1$. \\
\cline{2-2}
& \begin{minted}{python}
ds.SelectMany(lambda event: Zip({'pT': event.Jet_pt,
                                 'eta': event.Jet_eta})\
                            .Where(lambda jet: abs(jet.eta) < 1)\
                            .Select(lambda jet: jet.pT))
\end{minted}
\\
\hline
\multirow{2}{*}{4} & Plot the missing $E_\textrm{T}$ of events that have at least two jets with $p_\textrm{T} > 40 \textrm{ GeV}$. \\
\cline{2-2}
& \begin{minted}{python}
ds.Where(lambda event: event.Jet_pt\
                       .Where(lambda pT: pT > 40)\
                       .Count() >= 2)\
.Select(lambda event: event.MET_pt)
\end{minted}
\\
\hline
\end{tabular}
\caption{Implemented benchmark tasks. Only the FuncADL query is provided rather than the plot, which is a histogram of the elements in the result. These plots can be seen in the GitHub repository.}
\label{table:implemented}
\end{table}
\end{centering}

\begin{centering}
\begin{table}[ht]
\renewcommand{\arraystretch}{1.3}
\begin{tabular}{|>{\centering\arraybackslash}m{2mm}|m{\textwidth-54mm}|m{38mm}|}
\hline
\# & ADL benchmark task & Reason not implemented \\
\hline\hline
5 & Plot the missing $E_\textrm{T}$ of events that have an opposite-sign muon pair with an invariant mass between 60 and 120 GeV. & Missing cross join and \mbox{invariant} mass \\
\hline
6 & Plot $p_\textrm{T}$ of the trijet system with the mass closest to 172.5 GeV in each event and plot the maximum $b$-tagging discriminant value among the jets in the triplet. & Missing cross join, \mbox{sorting}, and invariant mass \\
\hline
7 & Plot the sum of $p_\textrm{T}$ of jets with $p_\textrm{T} > 30 \textrm{ GeV}$ that are not within 0.4 in $\Delta R$ of any lepton with $p_\textrm{T} > 10 \textrm{ GeV}$. & Missing cross join \\
\hline
8 & For events with at least three leptons and a same-flavor opposite-sign lepton pair, find the same-flavor opposite-sign lepton pair with the mass closest to 91.2 GeV and plot the transverse mass of the missing \mbox{energy} and the leading other lepton. & Missing cross join, sorting, and invariant mass \\
\hline
\end{tabular}
\caption{Unimplemented benchmark tasks. The features missing from FuncADL that are needed for full implementation of each task are indicated.}
\label{table:unimplemented}
\end{table}
\end{centering}

\clearpage

\section{Discussion}
\label{discussion}

As shown in the previous section, the data selections necessary for the first four ADL benchmark tasks were successfully implemented in FuncADL using the Uproot backend.
This demonstrates how the query language can express simple projections of sequences of events and object collections within each event, filtering of these sequences, and aggregation across elements.
These operations already encompass many basic data query use cases.

The selections required for the last four ADL benchmark tasks could not be expressed entirely within FuncADL, which indicates potential areas for improvement.
For example, the ability to sort a sequence by key values is an important query language feature that is missing.
Simple cross joins are in principle already well defined in the FuncADL syntax, but limitations in the Uproot backend do not yet support these.
The last feature needed for full implementation of the ADL benchmark tasks is the calculation of invariant mass.
The full details of the calculation could be written directly in the FuncADL query, but this would be inconvenient and error-prone.
As this is a common domain-specific function for HEP analyses, a more satisfactory solution would be to natively support four-vector operations.
All of these mentioned features are planned to be supported by future versions of FuncADL.

There are some limitations of FuncADL that are instilled by design.
As a query language, the functionality is limited to data extraction and transformation.
FuncADL is not intended to be used to run an entire analysis from beginning to end.
Calculating and applying scale factors, systematic variations, and creating actual histograms from selected data are examples of analysis tasks that are not well suited to FuncADL.
However, FuncADL is intended to be highly modular, such that it provides a uniform interface for querying data, either manually or via other analysis software.
This can reduce the burden on other analysis software that can instead focus on driving high-level analysis tasks.

Maintaining this narrow scope and modularity for FuncADL allows it to remain flexible and generalize well.
The xAOD backend is not preconfigured to use any particular collections or object properties, so it can be used with any xAOD file compatible with the same AnalysisBase release.
Similarly, the Uproot backend is not tied to any particular format of \texttt{TTree}.
This still allows for any package to be built on top of FuncADL that does impose a particular schema for an analysis user's convenience.

\section{Conclusions}
\label{conclusions}

The FuncADL project provides a declarative query-based interface to specify, retrieve, and transform data needed for HEP analyses.
This interface is not dependent on the underlying file format or data structures.
The query interface is identical whether executing the selection locally or remotely, either on a single machine or a cluster.
A subset of the query functionality has been demonstrated on the ADL benchmark tasks.
Some limitations have been identified, but these can currently be compensated for by combined usage with other packages and can be implemented directly by future extensions to the language.

\section{Acknowledgements}
\label{acknowledgements}

This work was supported by the National Science Foundation under Cooperative Agreement OAC-1836650.

\bibliography{main.bib}

\end{document}